\documentclass[aps,prl,twocolumn]{revtex4} 
\usepackage{graphicx}% Include figure files

\begin{document}
\title{Control of lasing in fully chaotic open microcavities by tailoring the shape factor}
\author{W. Fang$^1$, G. S. Solomon$^2$, and H. Cao$^1$}
 
\address{$^1$ Department of Physics and Astronomy, Northwestern University, Evanston, IL 60208. \\
$^2$ Solid-State Photonics Laboratory, Stanford University, Stanford, CA  94305; \\ and Atomic Physics Division, NIST, Gaithersburg, MD 20899-8423. } 

\begin{abstract}
We demonstrate experimentally that lasing in a semiconductor microstadium can be optimized by controlling its shape. Under spatially uniform optical pumping, the first lasing mode in a GaAs microstadium with large major-to-minor-axis ratio usually corresponds to a high-quality scar mode consisting of several unstable periodic orbits. Interference of waves propagating along the constituent orbits may minimize light leakage at particular major-to-minor-axis ratio. By making stadium of the optimum shape, we are able to maximize the mode quality factor and align the mode frequency to the peak of the gain spectrum, thus minimizing the lasing threshold. This work opens the door to control chaotic microcavity lasers by tailoring the shape factor.  
\end{abstract}

\maketitle

The Bunimovich's stadium billiard is a well-known model for classical and quantum chaos \cite{gutzwiller}. The ray mechanics in a stadium billiard exhibits ``full chaos'', i.e., there are no stable periodic orbits. However, a dense set of unstable periodic orbits (UPOs) are embedded in the sea of chaotic orbits. Although the UPOs are found with zero probability in the classical dynamics, in wave (quantum) mechanics they manifest themselves in the eigenstates of the system. There exist extra and unexpected concentrations, the so-called scars, of eigenstate density near UPOs \cite{heller}. Detailed studies have been carried out on closed or almost closed stadium cavities. A dielectric stadium, however, has its entire boundary open so that refractive escape and tunneling escape of light could happen at any point on the boundary. When optical gain is introduced to the stadium, light amplification may compensate the escape loss, leading to lasing action. In the past few years, lasing was realized in both scar modes and chaotic modes of semiconductor stadiums with certain major-to-minor-axis ratio \cite{haraPRE,haraPRL}. Highly directional output of laser emission was predicted \cite{schwefel} and confirmed in polymer stadiums \cite{Leb}. However, not only output directionality but also low lasing threshold is required for many applications of microlasers.

In this letter, we demonstrate experimentally that low lasing threshold can be obtained in a semiconductor microstadium by controlling its shape. Contrary to common expectation, modes of such a completely open fully chaotic microcavity may have long lifetime. These special modes are typically scar modes \cite{fang,lee}. When such a mode consists of several UPOs, the interference of partial waves propagating along the constituent orbits may minimize light leakage at certain major-to-minor-axis ratio \cite{fang}. Thus by tailoring the stadium shape, we are able to achieve optimum light confinement in a dielectric microstadium and thus a low lasing threshold. 

The sample was grown on a GaAs substrate by molecular beam epitaxy. The layer structure consists of  500nm AlGaAs and 200nm GaAs. In the middle of the GaAs layer there is an InAs quantum well (QW) of 0.6nm. The lower refractive index of AlGaAs layer leads to the formation of a slab waveguide in the top GaAs layer. Stadium-shaped cylinders were fabricated by photolithography and wet chemical etching. The major-to-minor-axis ratio of the stadiums was varied over a wide range while the stadium area remains nearly constant. The deformation of the stadium is defined as $\epsilon \equiv a/r$, where $2a$ is the length of the straight segments connecting the two half circles of radius $r$. 

To study their lasing properties, the stadium microcavities were cooled to 10K in a cryostat, and optically pumped by a mode-locked Ti-sapphire laser at 790nm. The pump beam was focused by an objective lens onto a single stadium. The emission was collected by the same lens, and sent to a spectrometer. As the pump power increased above threshold, certain peak corresponding to cavity resonance showed drastical increasing of peak intensity, accompanied with width narrowing. The threshold behavior indicated lasing phenomena. Lasing was realized in most stadiums with $\epsilon$ ranging from 0.4 to 2.2 and area $\sim 70\mu$m$^2$. Figure 1(a) shows the emission spectra of twelve stadiums slightly above their lasing thresholds so that we mainly see the first lasing mode. As $\epsilon$ increases, the first lasing mode jumps back and forth within the gain spectrum of the InAs QW. It is not always located near the peak of the gain spectrum.  At some deformation, e.g. $\epsilon$ = 0.94, 1.9, the first lasing mode is far from the gain maximum at $\lambda \sim$ 857nm. This phenomenon is not caused by lack of cavity modes near the maximum of the gain spectrum. A few small and broad peaks in the emission spectrum between 847nm and 857nm are due to cavity resonances. These resonances experiences higher gain than the lasing mode at $\lambda \approx 847$nm. The only reason they do not lase is their quality ($Q$) factors are low. This result indicates the lasing modes, especially the first one, must be high-$Q$ modes. However, when the lasing mode is away from the maximum of the gain spectrum, the relatively low optical gain at the lasing frequency results in high lasing threshold. This is confirmed in Fig. 1(b), which shows the lasing threshold strongly depends on the spectral distance between the first lasing mode and the maximum of the gain spectrum. Unlike many deformed microcavities \cite{nockel,narimanov}, the lasing threshold in a microstadium does not increase monotonically with the deformation, e.g., the lasing thresholds in stadiums of $\epsilon$ = 0.7 and 2.2 are nearly the same despite of their dramatically different deformations.

\begin{figure}
\includegraphics[width=7cm]{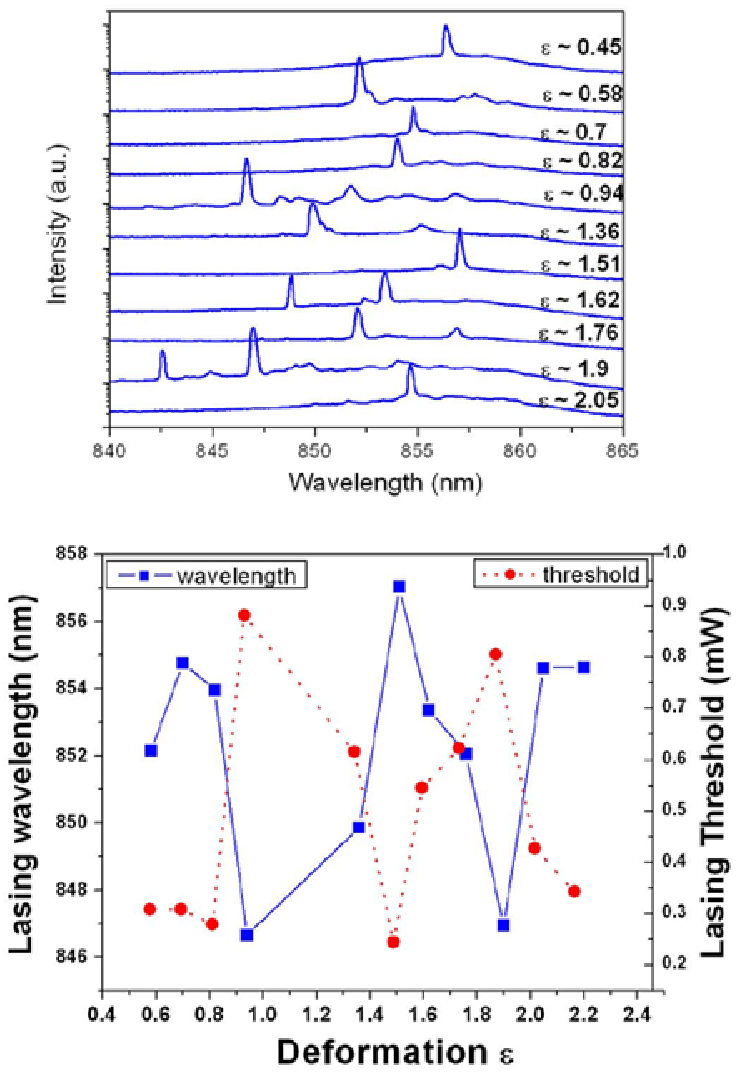}
\caption{\label{fig1}(a) Lasing spectra from twelve GaAs stadiums with different deformations. (b) Wavelength and lasing threshold of the first lasing mode as a function of $\epsilon$.}
\end{figure}

To investigate individual lasing modes in microstadiums, we used a narrow band pass filter to select one lasing mode and took its near-field image with a CCD camera. Figure 2 shows the measurement result of a stadium with $\epsilon = 1.51$. The solid curve is the emission spectrum when the narrow bandpass filter is tuned to the first lasing mode at $\lambda$ = 856.95nm. The near-field image  exhibits four bright spots on the curved part of  stadium boundary. We believe these four spots represent the positions of major escape of laser light from the stadium. They can be seen from the top because of optical scattering at the boundary. However, the scattering inside the stadium is so weak that the spatial intensity distribution of lasing mode across the stadium could not be observed from the top. By tuning the bandpass filter away from cavity resonances, we took the near-field image of amplified spontaneous emission (ASE) shown as the inset B of Fig. 2. The virtually constant intensity along the curved boundary suggests the ASE leaves the stadium mainly through the boundary of half circles instead of the straight segments. The clear difference between the near-field images of lasing mode and ASE not only confirms the bright spots in the former are from the laser emission, but also reveals the escape routes for laser emission and ASE are distinct. 

\begin{figure}
\includegraphics[width=7cm]{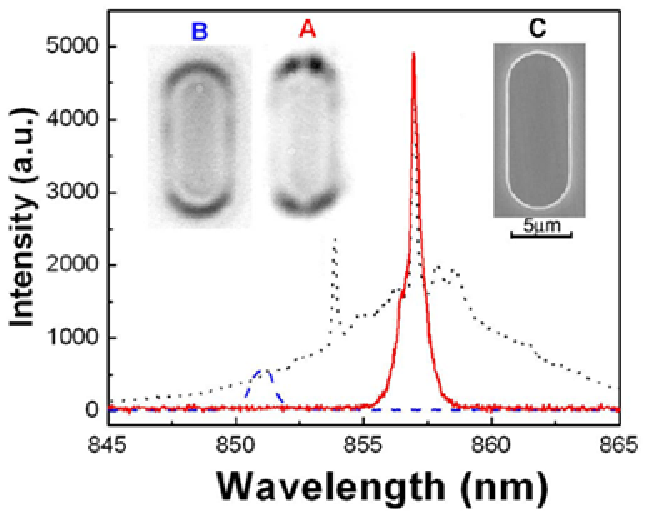}
\caption{\label{fig2} The dotted curve is the lasing spectrum from a GaAs stadium with $\epsilon$ =1.51. A bandpass filter of 1nm bandwidth selects the first lasing mode at 856.95 nm (solid curve), and the inset A is the corresponding near-field image taken simultaneously. The dashed curve is the spectrum when the bandpass filter is tuned away from cavity resonances, the corresponding near-field image of ASE is shown in the inset B. Top-view SEM image of the stadium is shown in the inset C.}
\end{figure}

To understand the experimental results, we simulated lasing in GaAs microstadiums. The polarization measurement of laser emission from the stadium side wall confirmed the lasing modes are transverse electric (TE) polarized (the electric field is parallel to the top surface of the stadium). From the calculation of TE wave guided in the GaAs layer, we obtained the effective index of refraction $n_{eff} \simeq 3.3$. The exact size and shape of the fabricated stadiums were extracted from the scanning electron microscope (SEM) images. Using the finite-difference time-domain (FDTD) method, we solved the Maxwell's equations for electromagnetic (EM) field inside and outside a two-dimensional (2D) stadium of refractive index $n_{eff}$ together with the four-level rate equations for electronic populations in the InAs QW \cite{nagra}. Light exiting the stadium into the surrounding air was absorbed by uniaxial perfectly matched layers. The external pumping rate for electronic populations was assumed uniform across the stadium, similar to the experimental situation. We gradually increased the pumping rate until one mode started lasing. Fourier transform of the EM field gave the frequency of the first lasing mode. Figure 3(a) shows the intensity distribution of the first lasing mode at $\lambda$ = 850.7 nm in the stadium with $\epsilon$ = 1.51 and area $\simeq 70 \mu$m$^2$. The pumping rate is slight above the lasing threshold. For comparison, we also calculated the high-Q modes in the passive stadium (without optical gain). Details of our numerical method can be found in Ref.\cite{fang}. By comparing the lasing mode with the resonant modes of the passive cavity, we find the first lasing mode corresponds to the highest-quality mode within the gain spectrum. As shown in Fig. 3(a), the spatial profile of the first lasing mode is almost identical to the mode at $\lambda$ = 850.7nm in the passive stadium. This result illustrates the nonlinear effect on the lasing mode is insignificant when the pumping rate is not far above the lasing threshold.  

\begin{figure}
\includegraphics[width=7cm]{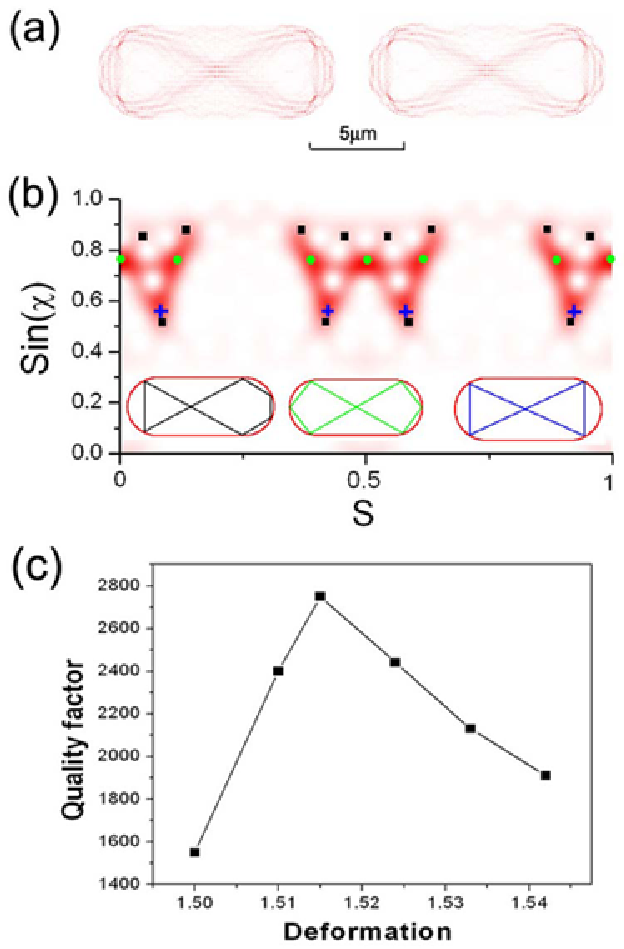}
\caption{\label{fig3}(a) Calculated intensity distribution of the first lasing mode in a stadium with $\epsilon$ = 1.51 (left), and the corresponding mode in the passive stadium without gain (right). Both modes have the wavelength 850.7 nm. (b) Husimi phase space projection of the mode in (a). The horizontal coordinate $s$ represents the length along the stadium boundary from the rightmost point, normalized by the stadium perimeter. The vertical axis corresponds to ${\rm sin} \chi$, where $\chi$ is the incident angle on the stadium boundary. The squares, dots and crosses mark the positions of three different types of UPOs shown in the inset. (c) Q-factor of the mode in (a) as a function of $\epsilon$.}
\end{figure}

The intensity of light escaping through the stadium boundary can be approximated by the intensity just outside the boundary. From the calculated lasing mode profile, we extracted the intensity  about 100nm outside the stadium boundary. To account for the finite spatial resolution in our experiment, the output intensity distribution along the stadium boundary was convoluted with the resolution function of our imaging system. The final result (dashed curve in Fig. 4) agrees well with the measured intensity along the stadium boundary (solid curve), especially since it reproduces the positions of four bright spots in the near-field image of the lasing mode. Since there is no other mode that has similar (low) lasing threshold and output intensity profile like the measured one, we conclude the first lasing mode observed experimentally in the stadium of $\epsilon$ = 1.51 corresponds to the calculated mode at $\lambda$ = 850.7nm. The slight difference (less than 1\%) in wavelength is within the experimental error of determining the refractive index of GaAs and AlGaAs at low temperature. The escape of ASE from a stadium is simulated by classical ray tracing in real space, as the interference effect can be neglected due to lack of coherence in ASE. Following the method in Ref.\cite{schwefel}, we calculated the intensity distribution of output rays along the boundary of a stadium with $n_{eff}$ = 3.3 and $\epsilon$ = 1.51 [dashed curve Fig. 4(b)]. The ray-tracing result agrees with the ASE intensity distribution obtained from the near-field image [solid curve in Fig. 4(b)].

\begin{figure}
\includegraphics[width=7cm]{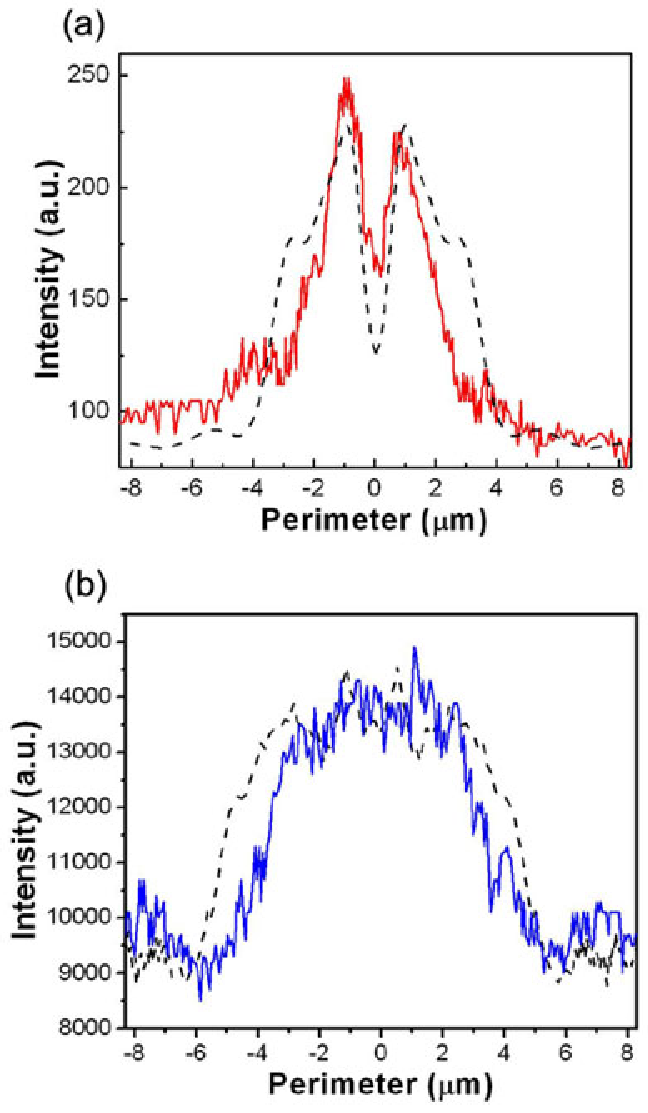}
\caption{\label{fig4}Output intensity of laser emission (a) and ASE (b) along the boundary of a GaAs stadium with $\epsilon$ = 1.51 and area $\simeq 70\mu$m$^2$. The range of the horizontal coordinate is half of the stadium boundary, from the center of one straight segment to the other. The solid curves are the experimental results extracted from the near-field images of the lasing mode and ASE in the insets of Fig. 3. The dashed curves are the numerical simulation results obtained with the FDTD method (a) and real-space ray tracing (b).}
\end{figure}

To find out the classical ray trajectories that the lasing modes correspond to, we obtained the quantum Poincar\'e sections of their wavefunctions. Figure 3(b) is the Husimi phase-space projection of the lasing mode in Fig. 3(a), calculated from its electric field at the stadium boundary. It reveals the lasing mode is a scar mode, and it consists mainly of three different types of UPOs plotted in the inset of Fig. 3(b). Since the constituent UPOs are above the critical line for total internal reflection, the lasing mode has long lifetime.  We calculated the quality factor of this mode in passive stadium as we varied the deformation $\epsilon$ around 1.51. Its $Q$ value first increases then decreases as $\epsilon$ increases, leading to a maximum at $\epsilon$ = 1.515. Such variation of quality factor is attributed to interference of waves propagating along the constituent UPOs \cite{fang,fuku}. The interference effect depends on the relative phase of waves traveling in different orbits. The phase delay along each orbit changes with the orbit length as $\epsilon$ varies. At some particular deformation, constructive interference may minimize light leakage out of the cavity, thus maximizing the quality factor. Since the actual deformation $\epsilon=1.51$ is nearly identical (within 0.3\%) to the optimum deformation ($\epsilon=1.515$), the mode is almost at the maximum of its quality factor. Furthermore, its frequency is close to the peak of the gain spectrum. Thus the lasing threshold is minimized, as shown in Fig. 1(b). 

% high-Q modes are multi-orbit scar modes, maximum Q at optimum deformation. design the stadium to be at the optimum deformation, and the also tune the wavelength to the peak of gain spectrum. it gives the lowest lasing threshold. In the case of spatially uniform pumping, the lasing threshold depends on the quality factor of the resonant mode and its spectral overlap with the gain spectrum. 

We simulated lasing in fabricated microstadiums with various deformations. By comparing the simulation results with the experimental data, we find the first lasing modes always correspond to high-quality scar modes of the passive cavities. This is because the gain spectrum of the InAs QW is broad enough to cover some of these modes. Note that not all the scar modes have high quality or long lifetime. Nevertheless the chaotic modes always have short lifetime, because their relatively uniform distributions in the phase space facilitate the refractive escape of light from the stadium.  If the gain spectrum is too narrow to contain any high-$Q$ scar modes, lasing may occur in low-$Q$ chaotic modes lying within the gain spectrum \cite{haraPRL} but with much higher threshold. 

One unique property of microstadium lasers is that the $Q$-spoiling is effectively stopped at large deformation \cite{fang}. High-$Q$ modes exist at large $\epsilon$, due to nonmonotonic change of quality factors of some multi-orbit scar modes with deformation. Indeed we observed different types of high-$Q$ scar modes consisting of several UPOs at various deformations in our simulation. This observation is supported by our experiment result that the lasing threshold at large deformation $\epsilon = 2.2$ is nearly the same as that at small deformation $\epsilon = 0.7$. Since the lasing wavelengths in these two stadiums of same area are nearly the same, the spectral overlap of the first lasing modes with the gain spectrum is almost identical. Therefore, the almost same lasing threshold implies the quality factor of the first lasing mode in the stadium with $\epsilon = 2.2$ is nearly identical to that with $\epsilon = 0.7$.  The effective stop of $Q$ spoiling does not exist in the elliptical cavity (an integral system) or the quadrupolar cavity (a partially chaotic system) \cite{nockel,narimanov}. Those systems exhibit continuous $Q$ spoiling with increasing deformation when the cavity area is fixed \cite{fang}. Thus a global increase of lasing threshold is expected when the major-to-minor-axis ratio of dielectric ellipse or quadrupole increases. 

In summary, we demonstrated experimentally that lasing in a semiconductor microstadium can be optimized by controlling its shape. By tuning the stadium shape to the optimum deformation, we not only optimize light confinement in the stadium but also extract the maximum gain by aligning the mode frequency to the peak of the gain spectrum. The simultaneous realization of the lowest cavity loss and the highest optical gain leads to minimum lasing threshold of a microstadium laser. As the dielectric microstadium represents a completely open fully chaotic cavity, this work opens the door to control chaotic microcavity lasers by tailoring its shape.  

We acknowledge Prof. Peter Braun and Dr. Gabriel Carlo for stimulating discussions. This work is supported by the MRSEC program of the National Science Foundation (DMR-00706097) at the Materials Science Center of Northwestern University.

\end{document}